\documentclass[aps,pre,onecolumn,notitlepage]{revtex4-2}

\usepackage[T1]{fontenc}
\usepackage[utf8]{inputenc}
\usepackage{lmodern}

\usepackage{amsmath,amssymb}
\usepackage{graphicx}
\usepackage{float}
\usepackage{multirow}
\usepackage{makecell}
\usepackage{tabularx}
\usepackage[table,xcdraw,dvipsnames]{xcolor}
\usepackage{longtable}
\usepackage{placeins}
\usepackage{booktabs}
\usepackage{siunitx}
\usepackage{microtype}
\microtypesetup{activate=true}

\begin{document}

\title{Geometric memory in incomplete phase transitions across dimensions}

\author{F.\ \c{T}olea}
\affiliation{National Institute of Materials Physics, POB MG-7, Bucharest-M\u{a}gurele, 077125, Romania}

\author{M.\ \c{T}olea}
\email{tzolea@infim.ro; tzolea123@yahoo.com}
\affiliation{National Institute of Materials Physics, POB MG-7, Bucharest-M\u{a}gurele, 077125, Romania}

\begin{abstract}
We model a \emph{direct} solid-state phase transition through a nucleation-and-growth process in which plates have simple, regular shapes---squares, cubes, or square-faced lamellae---and grow homothetically (self-similarly) until they either reach a randomly assigned maximum size or are stopped by impingement with previously formed plates. The \emph{reverse} transformation is represented by the preferential disappearance of smaller plates, while larger plates are retained during an incomplete reversion. A subsequent direct transformation therefore produces a modified plate-size distribution---a \emph{memory effect} that forms the main focus of this study. Building upon an earlier two-dimensional ($2\mathrm{D}$) formulation, we extend the model to \emph{cubes} ($3\mathrm{D}$) and to \emph{lamellar plates} ($3\mathrm{D}_{\!L}$) in order to examine how dimensionality affects transformation memory. We introduce a quantitative descriptor of memory, the \emph{Size Mass Ratio} (SMR), and find that memory is robust in all geometries but overall stronger in $2\mathrm{D}$ than in $3\mathrm{D}$ or $3\mathrm{D}_{\!L}$. We provide growth snapshots, arrest--regrowth cycles, size distributions, and DSC simulations, and we compute the Shannon size--entropy to quantify configurational diversity. Although motivated by the thermal memory effect in shape-memory alloys, the model more generally identifies a purely geometric mechanism for memory in first-order solid--solid transformations, highlighting the role of dimensionality and geometric blocking in controlling the strength of transformation memory.
\end{abstract}

\keywords{Phase-transition simulations; Statistical geometry; Thermal memory; Martensitic transformation}

\maketitle

\section{Introduction}

The theoretical description of solid--solid phase transitions spans from macroscopic thermodynamics to microscopic statistical models. Landau’s phenomenological theory \cite{Landau1937,LLStatPhys} introduced the order parameter concept and showed how the free energy landscape determines whether a transition is first or second order. Phase-field and Ginzburg--Landau-type approaches \cite{CahnHilliard1958,Khachaturyan1983} generalize this picture to spatially varying order parameters and naturally include interfaces, gradient effects, and microstructure evolution (with kinetic extensions such as Allen--Cahn). Complementary to these continuum perspectives, first-order kinetics in terms of stochastic nucleation and interface-controlled growth lead to the well-known Avrami temporal laws \cite{Avrami1939}. At a more microscopic level, lattice models such as the Ising, Potts, and Blume--Emery--Griffiths models \cite{Binder1987,Krumhansl1989} capture criticality, nucleation barriers, and cooperative behavior directly from nearest-neighbor interactions. Finally, for ferroelastic and martensitic transitions, strain-based phenomenological models \cite{Lookman2003} express the order parameter in symmetry-adapted strain components and link continuum elasticity with mesoscopic microstructure.

While the classical frameworks have been extremely successful in establishing thermodynamic and kinetic foundations, they remain largely continuum or mean-field in nature. They do not usually resolve, in a direct manner, the geometric constraints that arise during growth, impingement, and finite-domain blocking, as is the case in martensitic transformations of shape memory alloys (SMA). SMA undergo a non-diffusive phase transition from a high-temperature, high-symmetry phase (austenite) to a low-temperature, low-symmetry phase (martensite) (see, e.g., \cite{Otsuka,Planes,Jani,Vasilev,Pasquale,Kaufmann,Zheludev,Umetsu,Nnamchi}). These alloys also exhibit a surprising temperature memory: the ability to recall temperatures where a previous phase change was halted. This so-called thermal memory effect (TME) \cite{Airoldi1,Mad-Scripta2,R-AACTA2,Cui,Liu,Tang,WANG_SM,Tolea2,Vidal2,ToleaJALCOM,VidalCrespoTA}---sometimes called ``thermal arrest''---is still not fully understood and has received less attention than the shape aspect. In short, if a sample in martensite is heated and the martensite$\to$austenite transformation is intentionally stopped at a certain temperature (the arrest temperature, AT), then cooled back to martensite and finally heated again through the whole transformation, the calorimetric signal during this last heating will show a small dip at a temperature close to that previous AT.

Despite earlier approaches, a direct geometric mechanism for memory remains largely unexplored. In particular, the influence of dimensionality and growth geometry on the strength of such memory effects has received little systematic attention. This work generalizes a statistical geometry model introduced for the $2\mathrm{D}$ case \cite{Tolea1,ToleaPRE}, originally developed to explain TME in SMA, to the three-dimensional ($3\mathrm{D}$) and quasi-three-dimensional lamellar ($3\mathrm{D}_{\!L}$) cases in a single coherent framework. A central goal of the present work is to understand how this purely geometric memory behaves when moving from an idealized $2\mathrm{D}$ setting to more realistic three-dimensional situations. Real martensitic microstructures are inherently three-dimensional and often form plate-like variants with a pronounced habit plane.

Within this model, the transformation is represented as random nucleation and homothetic finite growth of regular units (squares in two dimensions, cubes in three dimensions, and lamellae in the quasi-three-dimensional case). Growth continues until direct contact with neighboring units produces a jammed configuration. During reverse transformations, smaller units are preferentially eliminated, producing a size-dependent hysteresis and a measurable memory effect in subsequent cycles. In all dimensionalities explored, this simple geometrical mechanism consistently produces memory of incomplete reverse transformations: the system ``remembering'' how far back it had been previously transformed.

Changes in the plate distribution after earlier incomplete cycles (i.e., the memory) are visualized through bar-chart size histograms and simulated calorimetric signals, and quantified using both a local size--mass ratio and a global Shannon size entropy. Our results indicate an overall stronger memory in $2\mathrm{D}$ than in $3\mathrm{D}$ and $3\mathrm{D}_{\!L}$.

The present model is intentionally minimal and focuses on the geometric aspects of nucleation, growth, and impingement. It does not include elastic interactions, long-range strain fields, or explicit energetic terms, which are known to play an important role in many real solid--solid transformations. In other words, under these strongly simplified assumptions, the memory emerges purely from geometry and geometric blocking of growth, offering a minimal mechanism by which memory can arise in first-order solid--solid transformations. Our goal is not to replace established continuum theories or other phenomenological descriptions of martensitic transformations (including stress-based approaches \cite{R-AACTA2}), but rather to complement them by providing an intuitive geometric perspective.

The outline of the paper is as follows.
Section~II introduces the adaptive nucleation--growth model in its $2\mathrm{D}$, $3\mathrm{D}$, and quasi-$3\mathrm{D}$ lamellar ($3\mathrm{D}_{\!L}$) forms.
Section~III examines memory effects under arrest--regrowth cycles.
Section~IV visualizes these effects by constructing simulated DSC signals for the reverse transformation, and Section~V quantifies them using the Shannon size entropy and the size--mass ratio (SMR).
Section~VI concludes the paper.

\section{Homothetic Nucleation--Growth Model for $2\mathrm{D}$, $3\mathrm{D}$ and $3\mathrm{D}_{\!L}$}
\label{sec:model}

\subsection{General Overview}

Although purely geometrical in nature, our model is inspired by the martensitic transformation in SMA, where martensite plates stop growing at finite dimensions and the transition proceeds primarily through the formation of new plates. Within the model, any newly formed plate is assumed to have an intrinsic maximum size it could, in principle, reach; however, this maximum is typically not attained in practice because pre-existing plates impose geometric constraints that arrest growth earlier. Naturally, such constraints become increasingly dominant as the transformation approaches completion (the jamming limit). More technically, the model represents a geometry-controlled transformation on a discrete lattice. Each transformed domain is a regular body: a square ($2\mathrm{D}$), a cube ($3\mathrm{D}$), or a thin plate constrained to a habit plane ($3\mathrm{D}_{\!L}$).

In more detail, the {\bf direct transformation} has the following features:

\begin{itemize}
    \item \textbf{Plate geometry and size range:}
    All plates have sizes within $[L_{\min},\, L_{\max}]$. Their shapes are regular: squares in $2\mathrm{D}$, cubes in $3\mathrm{D}$, and square--faced parallelepipeds in the $3\mathrm{D}_{\!L}$ model.

    \item \textbf{Nucleation (seeding):}
    Each new plate (``nucleus'' or ``seed'') is introduced at the minimum size $L_{\min}$, placed at a location chosen uniformly at random from all currently available positions.

    \item \textbf{Intrinsic maximum size and homothetic growth:}
    Every plate is assigned an intrinsic maximum size $L \in [L_{\min},\, L_{\max}]$. It then attempts to grow homothetically (self--similarly) toward this size.

    \item \textbf{Growth limitation by impingement:}
    Growth may stop earlier, at some $L_1 < L$, if further expansion is blocked in all directions by existing plates or by the system boundaries.

    \item \textbf{Termination of the direct transformation:}
    The direct phase transition ends once no empty region remains where a new seed of size $L_{\min}$ can be placed.
\end{itemize}

Note that no explicit energetic terms are used; the evolution is entirely geometric and governed by excluded volume (i.e., no overlap between plates is allowed).

The {\bf reverse transformation} is much simpler to describe and essentially has {\it one single feature}: it is modeled as the disappearance of plates in reverse size order---the smaller plates vanish first. Even though we do not introduce explicit thermodynamics here, the underlying justification is that smaller plates have a higher surface-to-volume ratio and are therefore the first to become unstable.

Let us briefly relate this assumption to two previous thermodynamic models addressing the thermal memory effect. In \cite{Tolea1}, the formation of martensite plates explicitly includes the energy penalty associated with the border length, which naturally leads to smaller plates transforming back first. Earlier, \cite{R-AACTA2} considered an even distribution of embedded stress among the martensite plates; in that framework, the reverse transformation begins with the plates having the highest stress density, which become unstable first as the temperature increases. Although based on different physical ingredients, the models \cite{Tolea1} and \cite{R-AACTA2} are thermodynamically similar in that they both impose an ordering in the reverse transformation.

In the present model we do not explicitly include temperature (in line with \cite{ToleaPRE}), but the reverse process can be interpreted as if an ``effective temperature''—monotonically related to the plate sizes—had to increase progressively for larger plates to transform back, corresponding to an athermal transformation.

To this end, we draw attention to the long-standing debate over whether the martensitic transformation (direct or reverse) is thermal or athermal, with arguments supporting both viewpoints for different alloys (see, e.g., \cite{Rao,Perez,Recarte,Vidal2}).

Recent work has explored the combination of physical modeling and statistical learning through energy-based neural-network formulations for mechanics problems \cite{Samaniego}. While the present study adopts a minimal geometric model, such approaches illustrate possible ways in which modeling frameworks might be combined with data-driven methods.

\subsection{Visualization of Phase-Transition Evolution up to Jamming}

Our simulations use a cubic/square discrete domain of edge $A$. For visualizations and clarity in this subsection we take $A=30$ (while for better statistics we later take $A=80$). Nucleation attempts are drawn uniformly among all currently available $L=L_{\min}$ seed positions. Each accepted nucleus receives an intrinsic target size $L\in (L_{\min},L_{\max})$ sampled uniformly. Growth proceeds by outward, shape-preserving dilation, accepting an increment only if all newly covered cells are empty; a domain becomes frozen when no further increment is feasible. A \emph{jammed} state is declared when no $L_{\min}^d$ seed fits anywhere. Ensemble statistics are computed over multiple independent realizations. In the simulations presented in this paper, we take $L_{\min}=3$, meaning, in particular, incomplete phase transitions because minimal seeds cannot fit into the remaining pores near jamming. Taking the limit value $L_{\min}=1$ corresponds to a complete phase transition, with the whole surface eventually transformed---such a situation is described in \cite{ToleaPRE} and does not bring qualitative differences for the $2\mathrm{D}$ case. $L_{\max}$ will be either 7 or 9 throughout the paper.

\begin{figure}[htbp]
  \centering
  \includegraphics[width=0.7\linewidth]{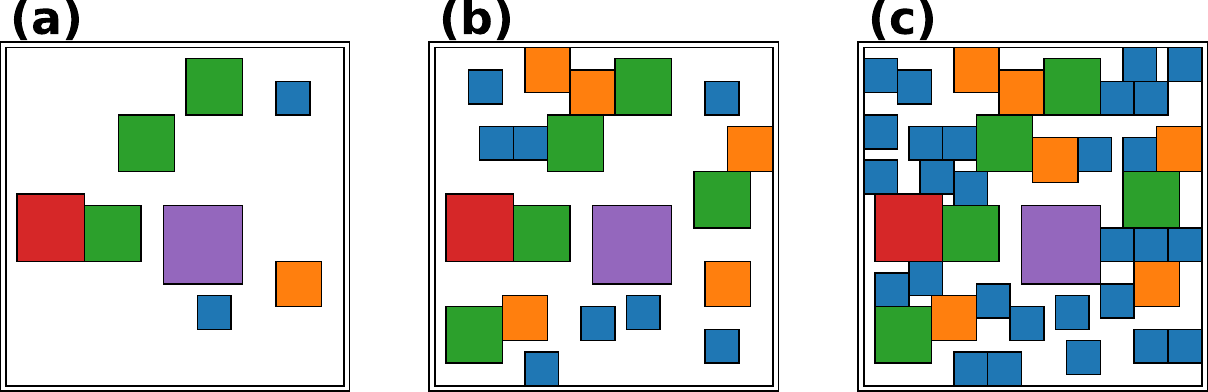}
  \caption{\textbf{$2\mathrm{D}$ squares.} Snapshots at (a) low coverage, (b) moderate coverage, and (c) the jamming limit.
  Visualization run for $A=30$.
Each new square is randomly placed among available $L_{\min}^2$ positions. $L_{\min}=3$ while $L_{\max}=7$ and the maximum intrinsic size $L$ for each new square is randomly chosen in this interval. The progressive crowding reduces available perimeter for growth and only small squares can grow towards jamming.}
  \label{fig:2D}
\end{figure}

\begin{figure}[htbp]
  \centering
  \includegraphics[width=0.75\linewidth]{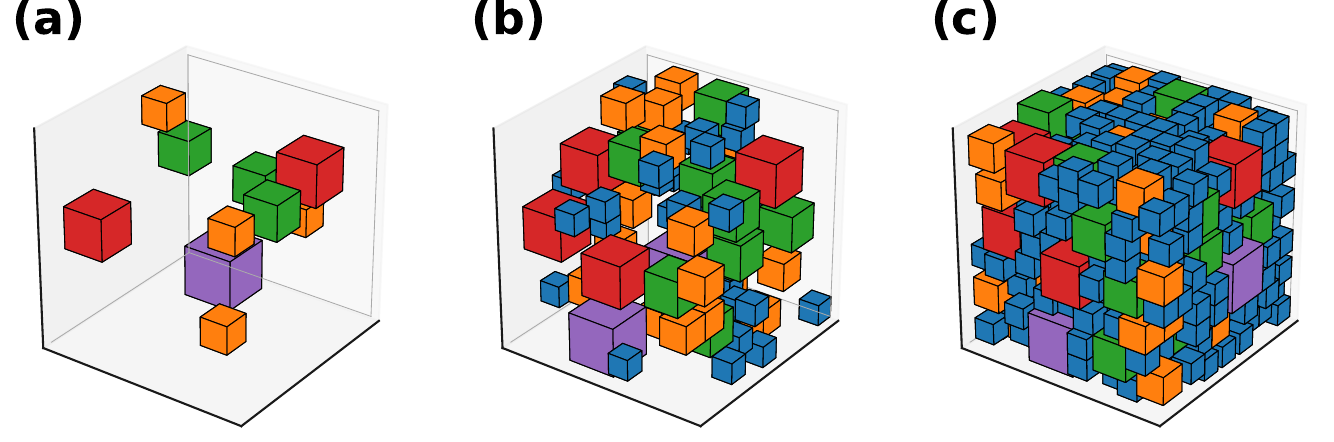}
  \caption{\textbf{$3\mathrm{D}$ cubes.} Snapshots at (a) low coverage, (b) moderate coverage, and (c) the jamming limit.
  Visualization run for $A=30$.
  Brute-force seeding uniformly among available $L_{\min}^3$ places, then growth up to a random maximum intrinsic size, not bigger than $L_{\max}=7$.
   In three dimensions, accessible free surface collapses faster than in $2\mathrm{D}$, favoring earlier lack of space for larger cubes.}
  \label{fig:3D}
\end{figure}

\begin{figure}[htbp]
  \centering
  \includegraphics[width=0.5\linewidth]{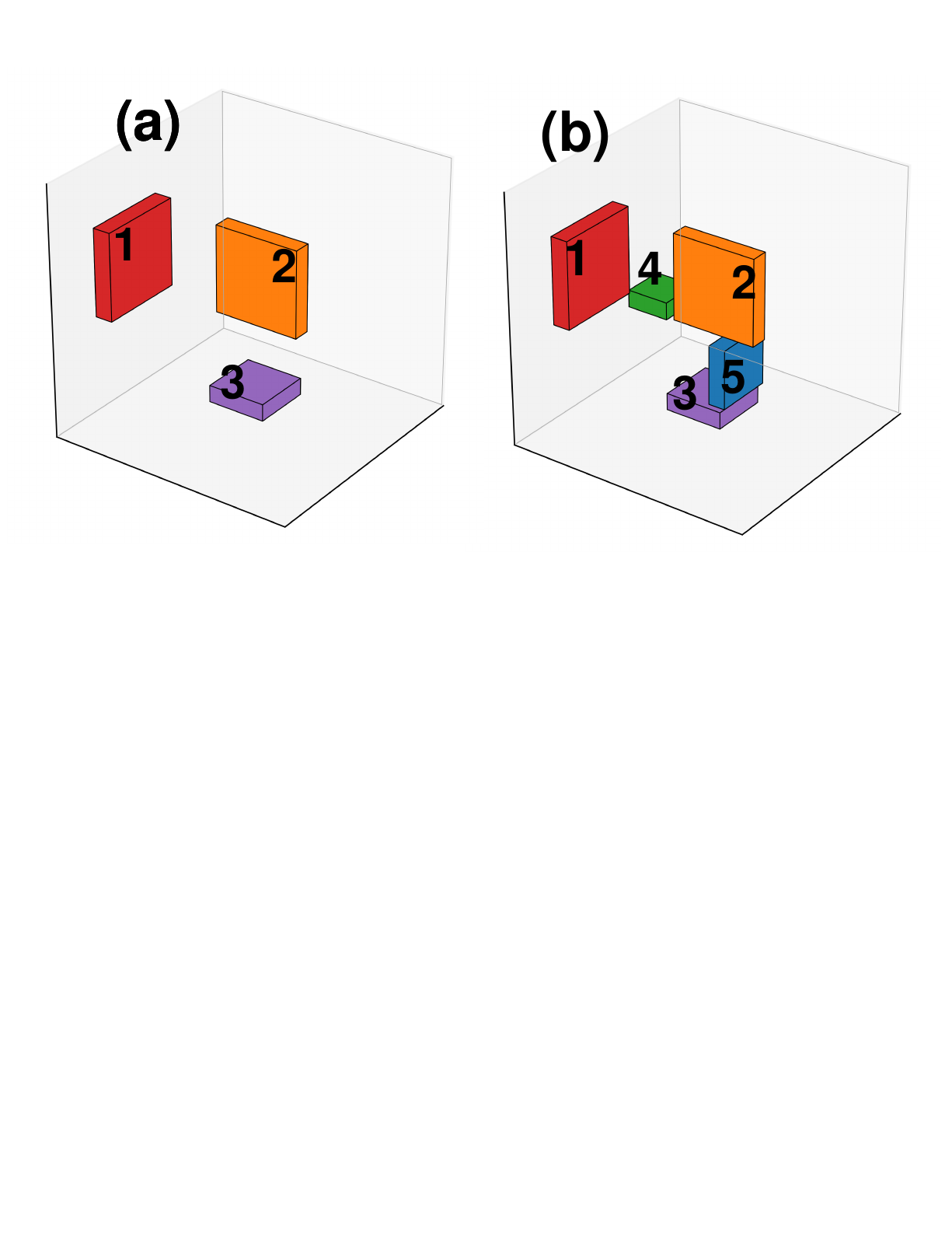}
  \vskip -7cm
  \caption{\textbf{$3\mathrm{D}_{\!L}$ lamellar plates.} (a) Three plates nucleate randomly and expand homothetically to their intrinsic sizes in the YZ, XZ, and XY planes. (b) As space becomes crowded, geometric constraints may prevent full growth, so plates 4 and 5 remain smaller due to impingement.}
  \label{fig:Manual}
\end{figure}

\begin{figure}[htbp]
  \centering
  \includegraphics[width=0.75\linewidth]{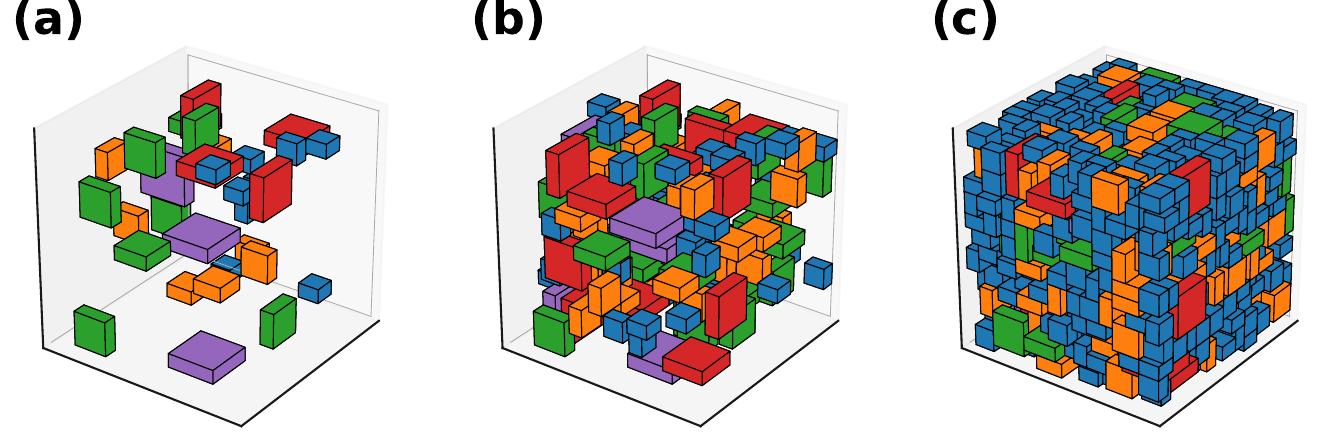}
  \caption{\textbf{$3\mathrm{D}_{\!L}$ lamellar plates.} Snapshots at (a) low coverage, (b) moderate coverage, and (c) the jamming limit.
  Visualization run for $A=30$.
  Perfect-square plates in XY/XZ/YZ planes with fixed thickness $TW=2$ along the normal.
  Brute-force seeding among available $L_{\min}^3$ seeds; $L_{\min}=3$, $L_{\max}=7$.
  Lateral impingement dominates and freezes many nuclei early, yielding a ``bottom-heavy'' size distribution.}
  \label{fig:3Dlam}
\end{figure}

Figures~\ref{fig:2D}--\ref{fig:3Dlam} provide representative snapshots at low coverage, intermediate coverage, and at the jamming limit for the three geometries considered in this paper: $2\mathrm{D}$, $3\mathrm{D}$, and $3\mathrm{D}_{\!L}$. Panels (a)--(c) in each figure therefore show how an initially sparse configuration of seeds evolves toward a crowded, highly partitioned microstructure, with colors indicating the plate sizes that will later appear as bins in the histograms of Sec.~III.

We stress again that when adding each new plate (a ``nucleus'' or ``seed'' of minimum size $L_{\min}$), we use brute-force seeding: all available positions are identified, and one of them is chosen at random. While this procedure is somewhat computationally expensive (particularly in $3\mathrm{D}$), it ensures that the final distributions do not rely on any particular choices of nucleation sites.

Fig.~\ref{fig:2D} illustrates the gradual filling of a large square domain (side $A=30$) with smaller squares whose sizes range from $L_{\min}=3$ to $L_{\max}=7$. In our model, each newly placed square nucleates at the minimum size and then ``grows’’ to a target size drawn randomly from the interval $(L_{\min},L_{\max})$, while preserving its square shape. Owing to this randomness in size selection, the early stages of the transformation contain comparable numbers of squares of all sizes. Fig.~\ref{fig:2D}a displays a single random placement of the first few squares, but when averaged over many realizations the number of squares per size tends to equalize. Panels (b) and (c) make visually clear how, as space fills, new squares are forced to stop at smaller $L$ and progressively occupy the remaining ``holes'', anticipating the quasi-uniform mass-per-size distribution at jamming.

As the transformation progresses, however, geometrical constraints become increasingly important. Many squares are forced to stop growing before reaching their assigned target size. Consequently, by the stage shown in Fig.~\ref{fig:2D}c, the smallest squares ($L=3$, shown in blue) already dominate in number. Upon averaging over a large ensemble, our model predicts that at the jamming limit the \emph{total surface area} (``mass'') contributed by each size becomes nearly uniform---despite the fact that the \emph{counts} of squares are nearly uniform only in the very early stages of the transformation. These statistical averages will be illustrated below, when the comparison with the regrowth after arrest is emphasized.

In $3\mathrm{D}$ (Fig.~\ref{fig:3D}), the side lengths are the same ($A=30$, $L_{\min}=3$, $L_{\max}=7$), and the nucleation--growth rules are unchanged, except that we now place cubes instead of squares. As in $2\mathrm{D}$, the initial stages contain cubes of all sizes due to the randomness of the size selection and the initially low incidence of geometrical blocking. However, as jamming is approached, the accessible free-growth volume shrinks more rapidly and biases the jammed state toward smaller cubes. As before, we show only a single random realization for visualization of the process; the statistical mass distributions obtained from many realizations will be presented in the next subsection. The visual impression from panels (b)--(c) is that large cubes become rare ``islands'' embedded in a sea of small ones, which is precisely the microscopic picture behind the biased histograms in Fig.~\ref{fig:Sizes3D}.

For the $2\mathrm{D}$ case, the detailed description of our model can be found in
\cite{Tolea1,ToleaPRE}, while the $3\mathrm{D}$ case is technically very similar.
The lamellar geometry ($3\mathrm{D}_{\!L}$) is slightly more complex, so we
describe it here in more detail. In the lamellar case, plates have fixed
thickness along the plane normal and grow only within their habit plane (XY,
XZ, or YZ). The width is taken as $TW = 2$, corresponding to two lattice sites
(one lattice constant), whereas $TW = 1$ would correspond to infinitely thin,
in--plane squares.

Figure~\ref{fig:Manual}(a) illustrates the possible orientations of lamellar
plates in the XY, XZ, and YZ planes. The spatial position, orientation, and
intrinsic maximum size (the square facet side restricted to the range
$[L_{\min}, L_{\max}]$) are chosen randomly. As the transformation proceeds, the
probability increases that the growth of new plates (labelled 4 and 5 in
Fig.~\ref{fig:Manual}(b)) is prematurely stopped due to geometric constraints
imposed by already existing plates.

Figure~\ref{fig:3Dlam} shows an initial stage, an intermediate stage, and the
jammed configuration for the $3\mathrm{D}_{\!L}$ lamellar geometry.

The $3\mathrm{D}_{\!L}$ case (Fig.~\ref{fig:3Dlam}) further exaggerates lateral blocking: plates expand only within their habit plane and therefore tend to freeze earlier, since they can be blocked both by plates lying in the same plane and by plates in perpendicular planes. This yields a size spectrum strongly tilted toward smaller $L$, even more pronounced than in the full $3\mathrm{D}$ (cubes) case. In practice, one can already see in panel (c) that the microstructure is dominated by short lamellae, with only a few large plates percolating across the sample, which anticipates the strong initial bias of Fig.~\ref{fig:Sizes3DLam}a.

We have been talking in this section about the plate-size distribution, which is a key element in our model. The statistics will be given in the next sections, to be directly compared with the distributions after arrested reverse transformations. The idea is that the snapshots in Figs.~\ref{fig:2D}--\ref{fig:3Dlam} provide the ``real-space'' counterpart of the one-dimensional histograms that will encode the memory effect.

To this end, we would like to make a brief comment. While the jamming limit is clearly defined, another conceptual quantity may also be relevant for describing such systems, namely the \emph{percolation threshold} (i.e., the average filling at which a system-size cluster of touching plates first appears). Intuitively, the percolation threshold should be smaller than the jamming limit, but mathematically even the jamming condition does not guarantee system-wide connectivity, since it only implies the disappearance of available nucleation sites. While this is likely an interesting issue for all geometries, we believe the 3D lamellar case to be particularly promising. This aspect was raised to us by one of the Referees, and we plan to address it in future work.

\section{Arrest--regrowth cycles and geometric memory}

As described above, a \emph{direct} transformation is simulated by nucleation followed by homothetic growth, continuing until a jammed state is reached (no additional $L_{\min}^d$ seed can be inserted).
A \emph{reverse} step is, however, not random, but modeled as a size-selective disappearance process in which the smallest domains vanish first.
Interrupting the reverse step at a chosen size (the ``arrest size’’ $s$) leaves a modified geometric configuration for the subsequent direct run.
This altered availability landscape, rather than any preferential nucleation rule, is the sole origin of the geometric memory.
Repeating the same arrest level on successive cycles (\emph{hammering}) reinforces the imprint at size $s$ by repeatedly reapplying the same geometric constraint.

\begin{figure}[htbp]
  \centering
  \includegraphics[width=0.6\linewidth]{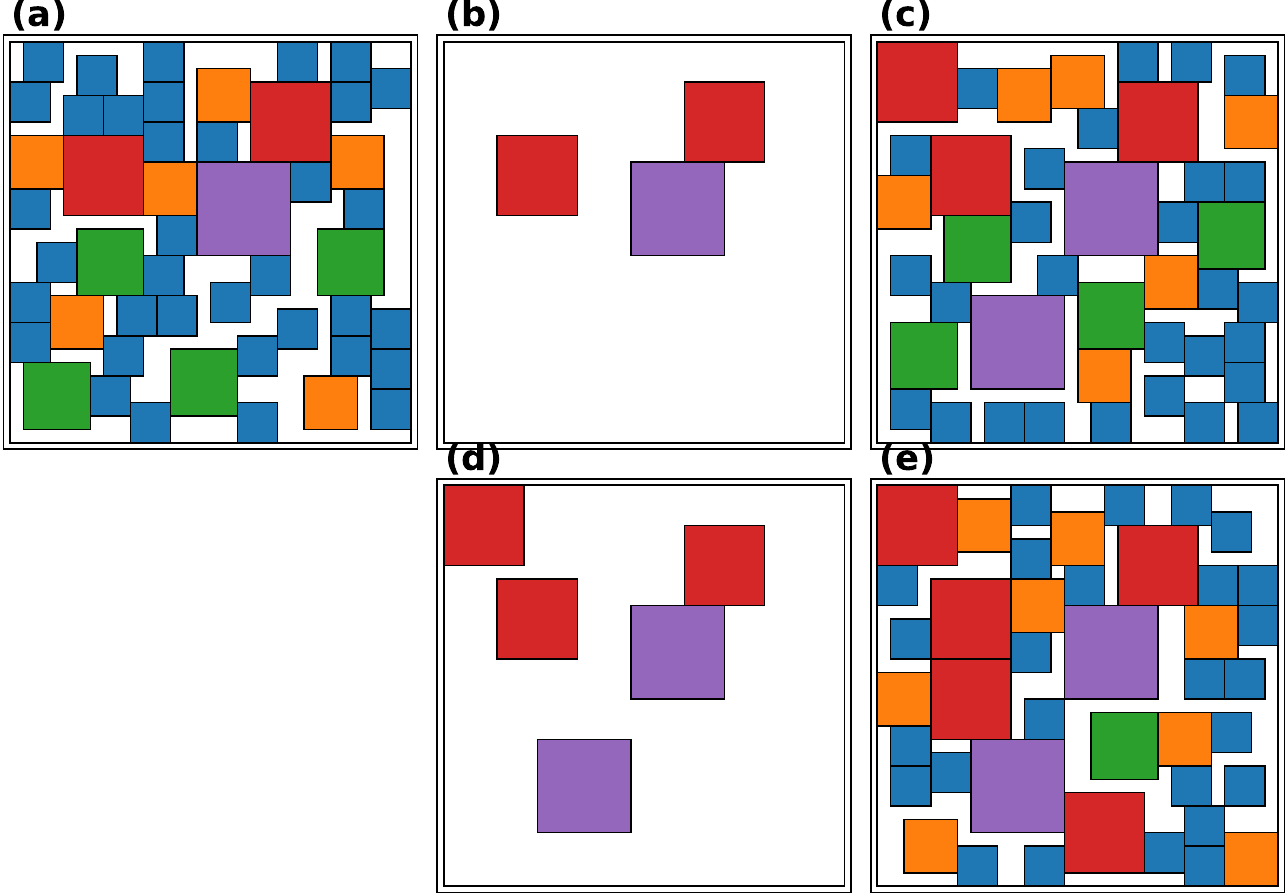}
  \caption{\textbf{$2\mathrm{D}$ squares.} Panels a)--e) correspond to the stages A--E (see description in text) of two successive arrest--regrowth cycles.
  (a) Initial fully jammed configuration after complete growth.
  (b) First arrest: only plates with linear size larger than the threshold $s$ are retained as seeds.
  (c) Jammed configuration after regrowth from these seeds.
  (d) Second arrest, applying the same threshold $s$ to the configuration in (c).
  (e) Final jammed configuration after the second regrowth. Repeating the same arrest (``hammering'') produces an increasingly sharp minimum in the size spectrum around $s$.}
  \label{fig:Fig2D_Regrow}
\end{figure}

\begin{figure}[htbp]
  \centering
  \includegraphics[width=0.7\linewidth]{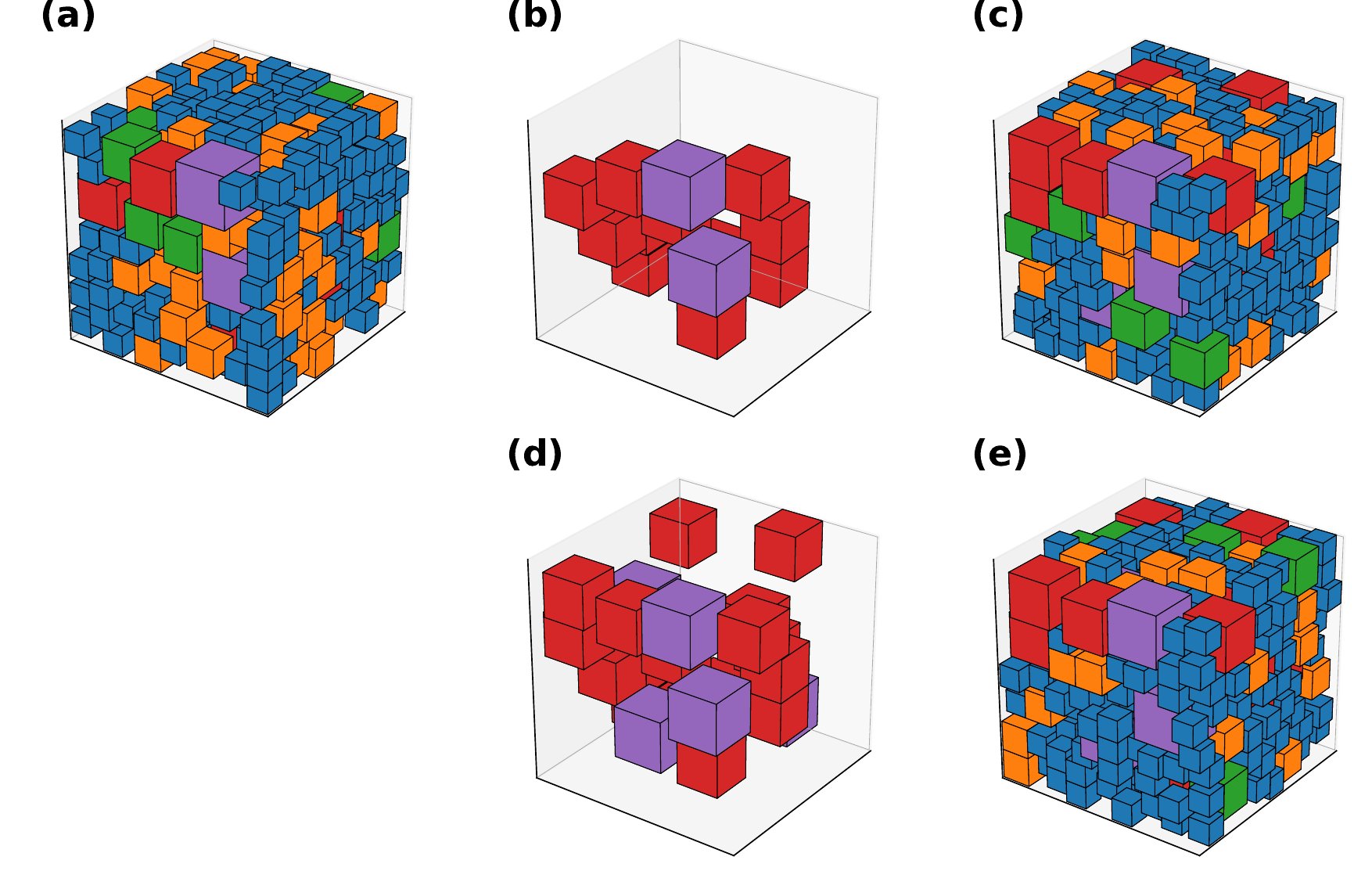}
  \caption{\textbf{$3\mathrm{D}$ cubes.} Panels a)--e) illustrate stages A--E (see description in text) of an initial full transformation followed by two successive arrest--regrowth cycles, in direct analogy with the $2\mathrm{D}$ case shown in Fig.~\ref{fig:Fig2D_Regrow}.
  Only the larger cubes (red/purple) persist through each arrest, while smaller cubes are removed and then re-nucleation and growth occur again until jamming.}
  \label{fig:Fig3D_Regrow}
\end{figure}

\begin{figure}[htbp]
  \centering
  \includegraphics[width=0.7\linewidth]{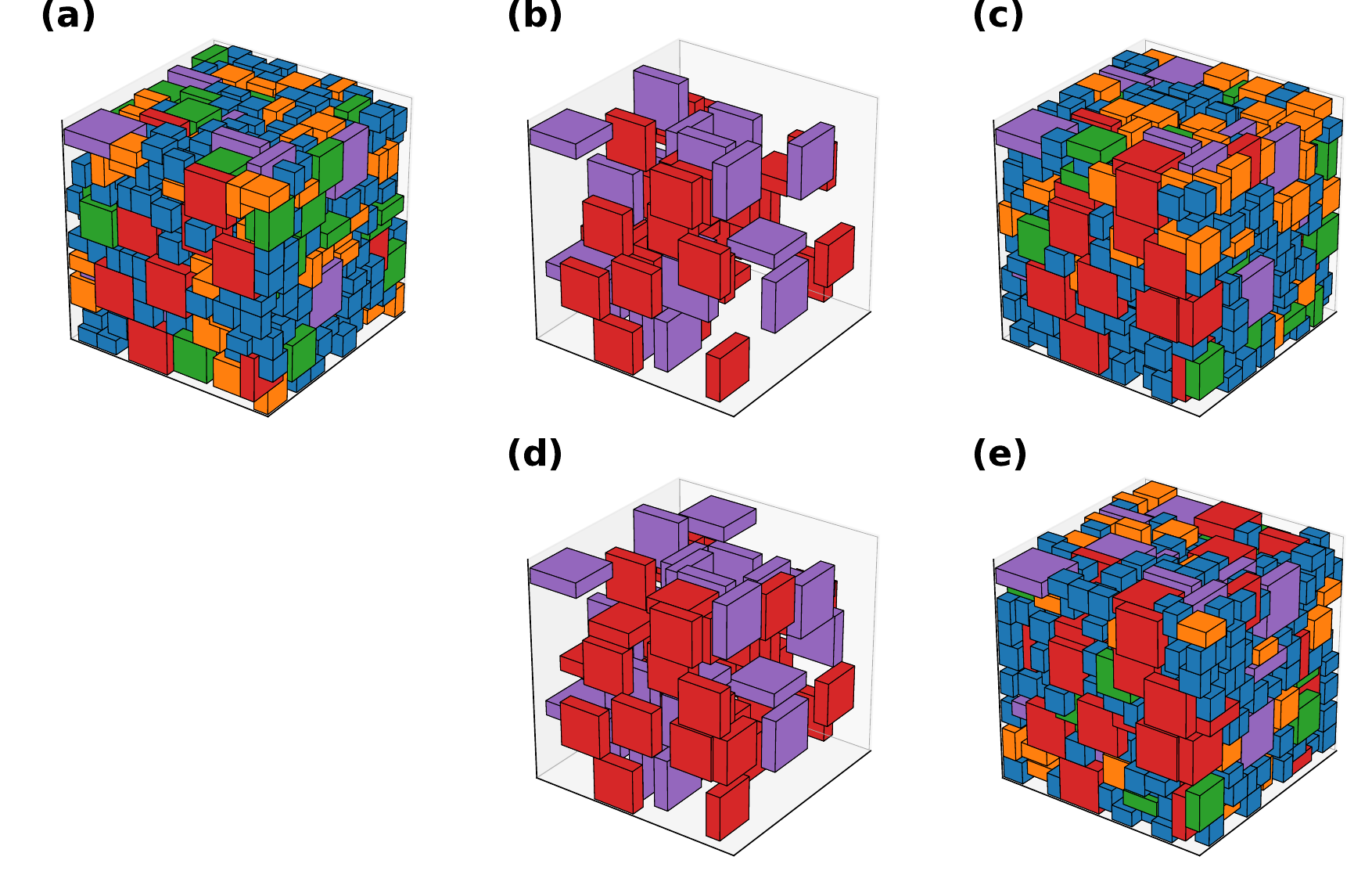}
  \caption{\textbf{$3\mathrm{D}_{\!L}$ lamellae.} Stages A--E of two successive arrest--regrowth cycles, analogous to the $2\mathrm{D}$ and $3\mathrm{D}$ cases shown in Figs.~\ref{fig:Fig2D_Regrow}--\ref{fig:Fig3D_Regrow}.
  Large (red/purple) lamellae survive both arrests, while smaller lamellae are removed at each arrest and re-nucleation and growth occur again until jamming.}
  \label{fig:Fig3DLam_Regrow}
\end{figure}

In this section we illustrate---both visually and through ensemble-averaged statistics---the complete arrest--regrowth--memory cycle.
The stages A--B--C--D--E correspond to the labels a), b), c), d), e) in Figs.~\ref{fig:Fig2D_Regrow}--\ref{fig:Fig3DLam_Regrow}, and denote, respectively:

\begin{itemize}
    \item {\bf A} refers to an initial complete (jammed) direct transformation started from scratch (with no pre-existing plates).
    \item {\bf B} refers to a partial reverse transformation, during which only the plates below a certain size disappear, while the larger ones are preserved.
    \item {\bf C} refers to the subsequent direct re-transformation, starting from the partially reversed state {\bf B}, during which new plates nucleate and grow within the geometrical constraints imposed by the preserved large plates.
    \item {\bf D} refers to an incomplete reverse transformation applied after {\bf C}, removing once more only plates below the same size threshold and leaving the larger ones intact.
    \item {\bf E} refers to the final direct re-transformation starting from the state {\bf D}, where regrowth proceeds under an even more constrained geometry due to the larger population of plates that remain untransformed.
\end{itemize}

\begin{figure}[htbp]
  \centering
  \includegraphics[width=0.7\linewidth]{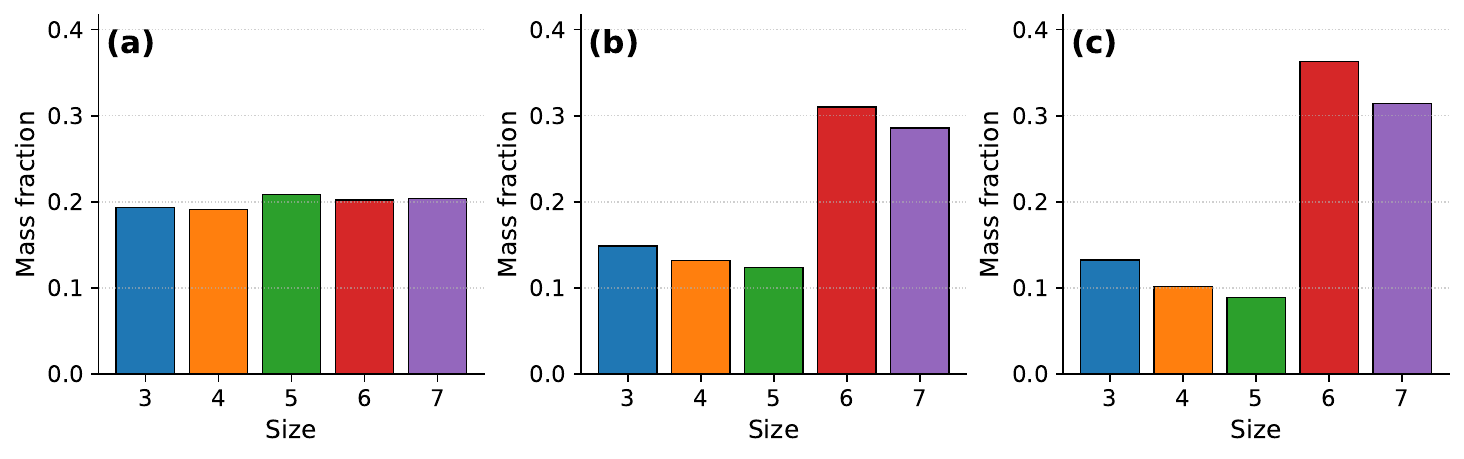}
  \caption{\textbf{$2\mathrm{D}$ squares.}
  Normalized mass distributions per size $L$ at stages
  \textbf{(a)}~A (initial jam),
  \textbf{(b)}~C (re-growth after first arrest), and
  \textbf{(c)}~E (re-growth after second, ``hammer'' arrest).
  Mass is $\sum L^{2}$; bars show ensemble averages.
  Progressive narrowing reflects elimination of small squares and preferential regrowth of larger ones. The arrested bin is a visible local minimum that deepens from C to E.}
  \label{fig:Sizes2D}
\end{figure}

Figures~\ref{fig:Fig2D_Regrow}--\ref{fig:Fig3DLam_Regrow} therefore implement two reverse arrests at the same physical size $s$, with a full regrowth to jam between arrests. In all geometries, the first arrest removes small domains and preserves larger ones (panels b), while the subsequent regrowth (panels c) inherits a modified availability landscape. Repeating the same arrest (panels d) leaves more ``big'' plates untransformed; the final regrowth (panels e) keeps a stronger imprint. The effect of plate-size distribution change is not very clear in the $3\mathrm{D}$ geometry, which is why the bar charts complement the picture. In other words, the eye easily follows how the red/purple plates percolate through the cycles, but the precise loss of intermediate sizes is better captured by the histograms discussed below.

\begin{figure}[htbp]
  \centering
  \includegraphics[width=0.7\linewidth]{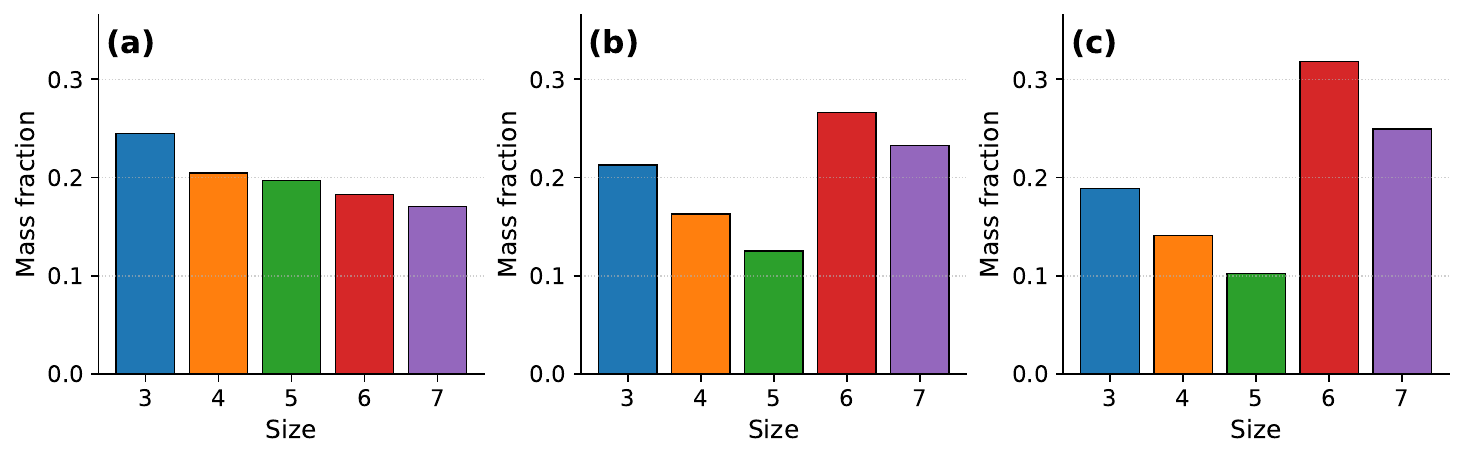}
  \caption{\textbf{$3\mathrm{D}$ cubes.}
  Normalized mass distributions per size $L$ at stages
  \textbf{(a)}~A,
  \textbf{(b)}~C, and
  \textbf{(c)}~E.
  Mass is $\sum L^{3}$; ensemble-averaged histograms.
  Removing small cubes at arrest shifts contribution toward larger sizes on re-growth. As in $2\mathrm{D}$, the arrested bin deepens under hammering.}
  \label{fig:Sizes3D}
\end{figure}

\begin{figure}[htbp]
  \centering
  \includegraphics[width=0.7\linewidth]{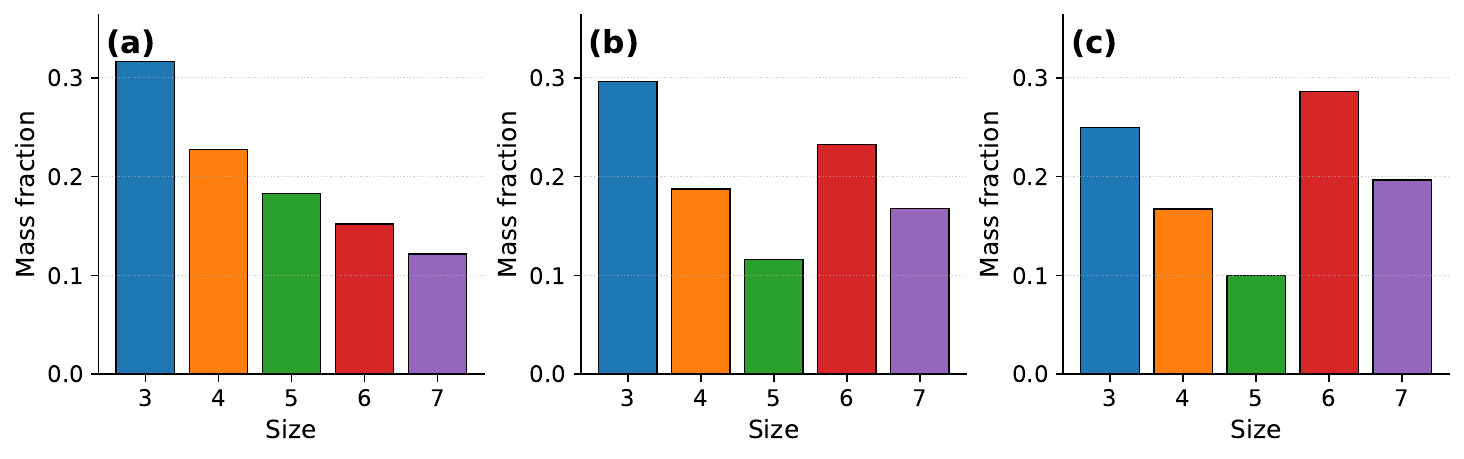}
  \caption{\textbf{$3\mathrm{D}_{\!L}$ lamellae.}
  Normalized mass distributions per in-plane edge $L$ at stages
  \textbf{(a)}~A,
  \textbf{(b)}~C, and
  \textbf{(c)}~E.
  Mass is $\sum (L^{2}\!\times\!TW)$ with fixed thickness $TW$. The panel (a) shows a markedly biased distribution after the initial jamming, while after a single (b) or double (c) arrest there is a clear depletion at the arrest size ``5''. }
  \label{fig:Sizes3DLam}
\end{figure}

To quantify these trends, Figs.~\ref{fig:Sizes2D}--\ref{fig:Sizes3DLam} report the mass-per-size histograms across A/C/E for each geometry. In all cases, the arrested bin becomes a local minimum whose depth increases under hammering. But it is also insightful to discuss the model’s prediction for the plate-size distribution after the initial jamming (before any arrest). The arrested size itself is indicated by the bin that is selectively depleted from A to C and then further from C to E, and this mirrors the visual reduction of plates of that characteristic color in Figs.~\ref{fig:Fig2D_Regrow}--\ref{fig:Fig3DLam_Regrow}.

For the $2\mathrm{D}$ case, the initial jamming leads to a quasi-even total area covered by squares of each size (the ``mass fraction''), as seen in Fig.~\ref{fig:Sizes2D}a. For the $3\mathrm{D}$ case, we observe a slightly biased distribution favoring smaller sizes, indicating an increased role of geometric constraints; see Fig.~\ref{fig:Sizes3D}a. The imbalance becomes even stronger in the $3\mathrm{D}_{\!L}$ case, where lamellae can be blocked both by in-plane plates and by perpendicular ones, as shown in Fig.~\ref{fig:Sizes3DLam}a. Panels (b) and (c) in each figure then display how the spectrum is reshaped by one and two identical arrests, respectively, offering a direct histogram-level view of the memory effect.

\section{Simulated DSC signal: visualizing the memory dip}

As mentioned, one aim of our model is to reproduce the TME in SMA, which is evidenced experimentally by calorimetric measurements (usually performed in a DSC, see, e.g. \cite{Airoldi1,Mad-Scripta2,R-AACTA2,Cui,Liu,Tang,WANG_SM,Vidal2,ToleaJALCOM}). The martensite--austenite phase transition is characterized by a peak in the DSC signal which, in the case of an arrested (incomplete) transformation, subsequently shows a specific dip or a ``shoulder'' near the arrest temperature. This is what we are trying to simulate in this section, starting from plate-size distributions calculated previously. In the reverse transformation, the plates ``disappear'' in reverse size order, so the DSC signal associated with each plate size should be proportional to the total area covered by that size (mass), as described in more detail below.

\begin{figure}[htbp]
  \centering
  \includegraphics[width=0.6\linewidth]{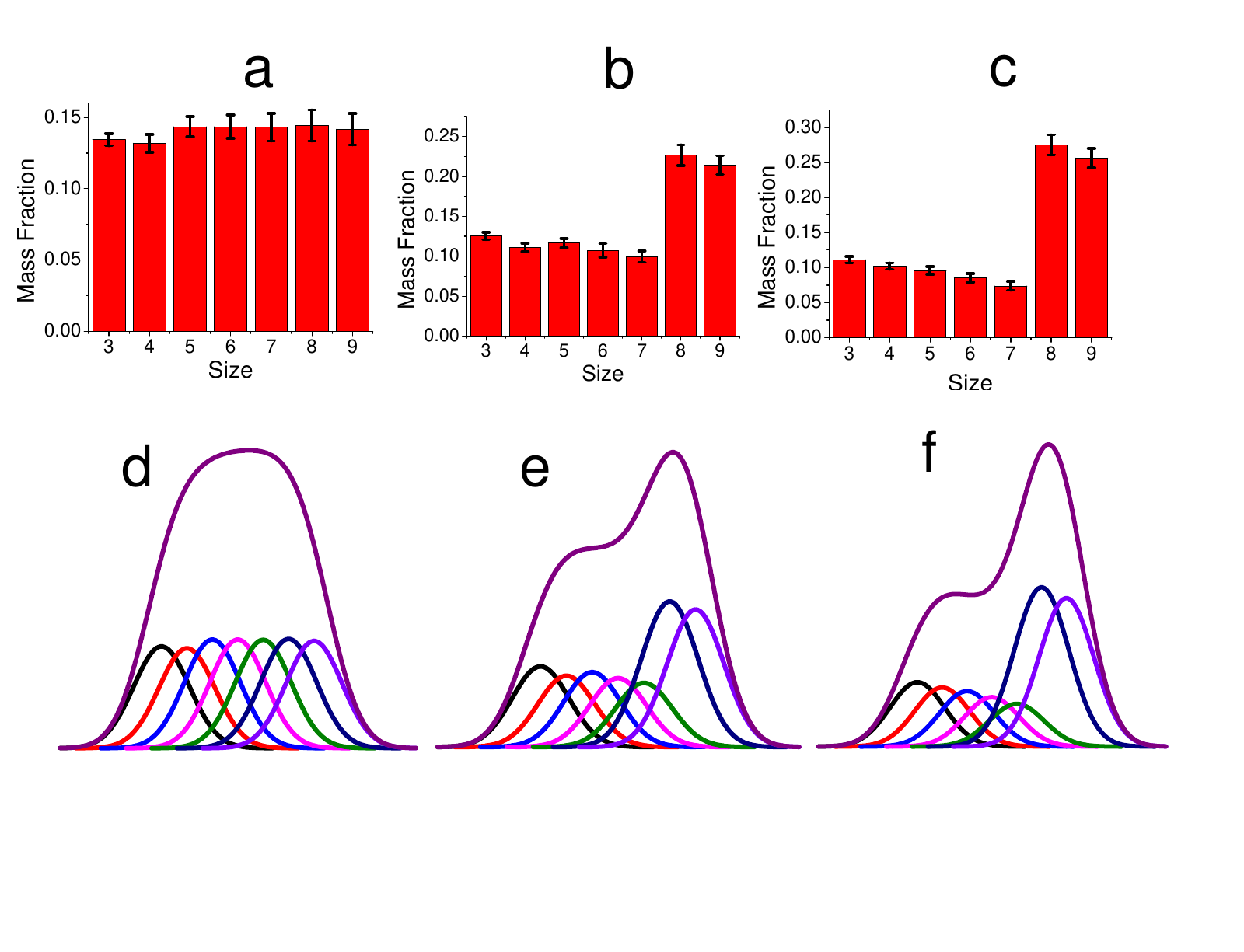}
  \vskip -1cm
  \caption{\textbf{$2\mathrm{D}$ squares.}
  Size--mass evolution across stages A and C/E, including the error bars (top row, panels a--c).
  The simulated reverse-DSC surrogate (bottom row, panels d--f) is constructed as described in the text.
  The dip (more accurately, a shoulder) near the prior arrest ``temperature'' in panel~e reflects the memory imprint, which becomes more pronounced under hammering, as seen in panel~f.
}
  \label{fig:2D_DSC}
\end{figure}

To visualize a DSC-like signal during the reverse transformation, each disappearing size bin $L$ is mapped to a nominal temperature $T(L)$, chosen as a monotone function of $L$ (larger domains $\mapsto$ higher $T$, consistent with later removal during reverse transformation).
A heat-flow surrogate $q(T)$ is then obtained by depositing the mass removed from bin $L$ into a narrow kernel (e.g., Gaussian) centered at $T(L)$, and summing contributions over all bins.
The kernel bandwidth affects only visual smoothness; all comparative features---especially the dip at the prior arrest temperature---are robust to this choice.

\begin{figure}[htbp]
  \centering
  \includegraphics[width=0.6\linewidth]{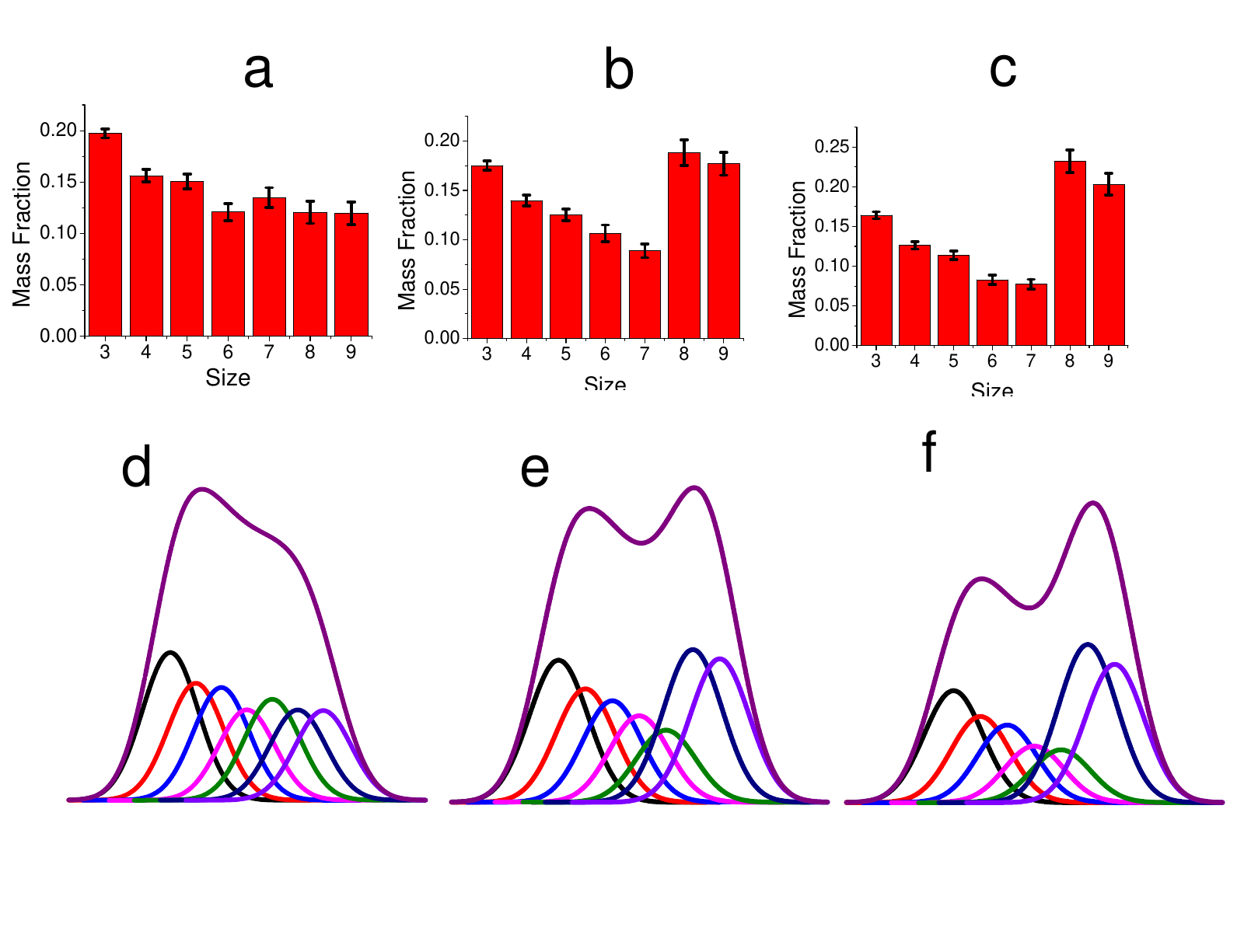}
  \vskip -1cm
 \caption{\textbf{$3\mathrm{D}$ cubes.}
  Panels~a--c: size--mass evolution across stages A/C/E, including the error bars.
  Panels~d--f: simulated reverse-DSC surrogate (see description in the text).
  The arrested bin generates a reproducible dip in the subsequent reverse run (panels~e and~f).
  The dip appears visually clearer than in $2\mathrm{D}$ because the initial distribution in panel~a already corresponds to a tilted DSC peak in panel~d.
}

  \label{fig:3D_DSC}
\end{figure}

\begin{figure}[htbp]
  \centering
  \includegraphics[width=0.6\linewidth]{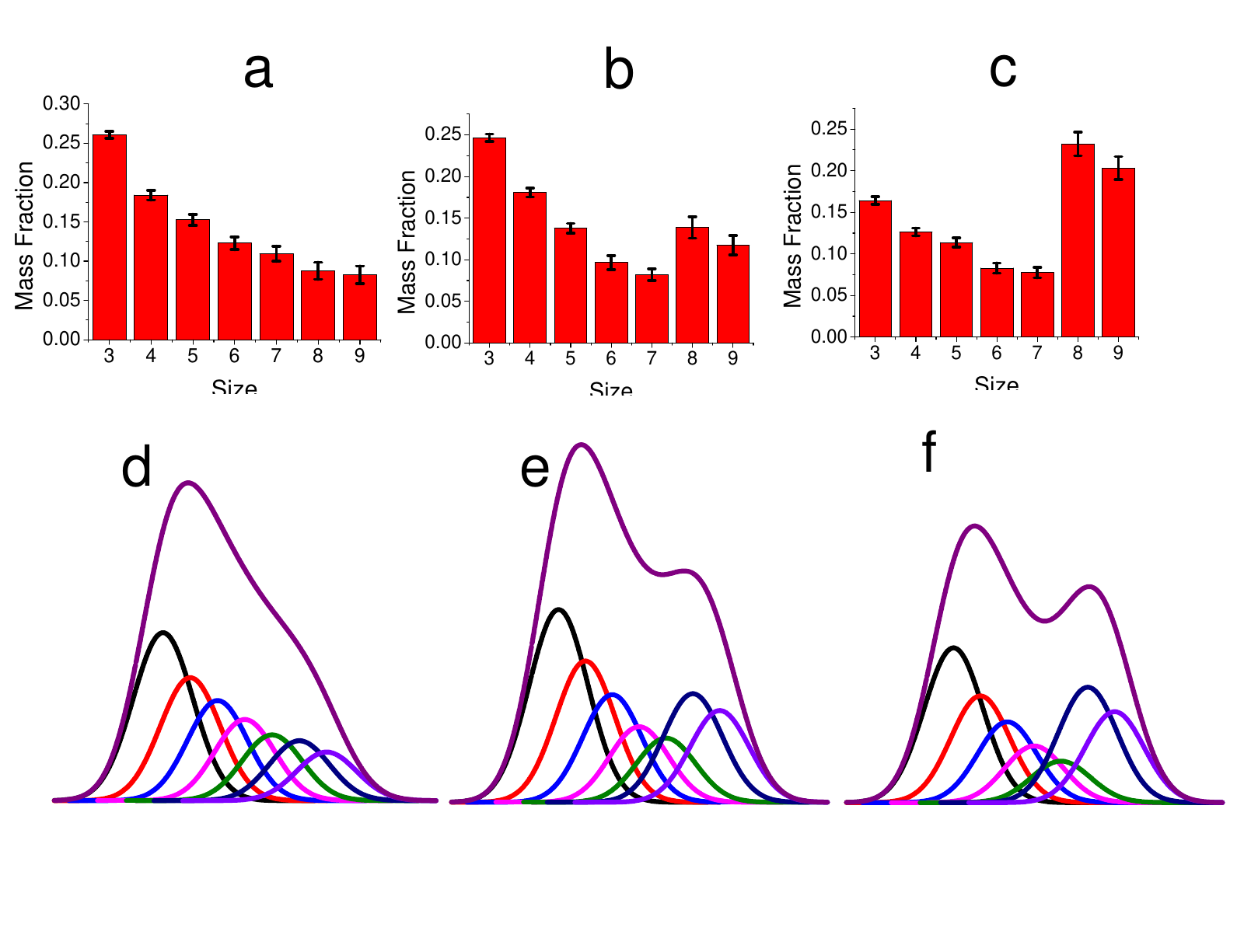}
  \vskip -1cm
\caption{\textbf{$3\mathrm{D}_{\!L}$ lamellae.}
  Panels~a--c: size--mass evolution across stages A/C/E, including the error bars.
  Panels~d--f: simulated reverse-DSC surrogate (see description in the text).
  The biased initial distribution in panel~a produces an asymmetric DSC peak in panel~d, while the distributions after arrest and regrowth show a ``shoulder'' after the first arrest in panel~e and a clearer modification after the hammer procedure, as reflected in panel~f.
}
  \label{fig:3DL_DSC}
\end{figure}

Figs.~\ref{fig:2D_DSC},~\ref{fig:3D_DSC},~\ref{fig:3DL_DSC} also include error bars on the size distribution (upper panels), which are sufficiently small that they do not affect the main conclusions regarding mass redistribution after arrest. Error bars represent the standard error of the mean (SEM), calculated as $\mathrm{SEM}=\sigma/\sqrt{N}$, where $\sigma$ is the standard deviation obtained from independent realizations and $N$ is the number of realizations.

This mapping translates the size distribution into a notional temperature scale and enables construction of a simulated reverse-DSC curve that qualitatively resembles experimental heat-flow traces.
In this representation, the depletion at the previously arrested size manifests as a dip or ``shoulder'' in the simulated DSC signal at the corresponding temperature.
Repeating the same arrest level (the ``hammer'' protocol) deepens this dip, reflecting reinforcement of the geometric memory. In practice, Figs.~\ref{fig:2D_DSC}--\ref{fig:3DL_DSC} should be read from top to bottom: the upper panels show which size bins disappear at each effective temperature step, while the lower panels show how this loss translates into the shape of the synthetic DSC peak.

For the numerical runs presented in this section (and for the averaged bar charts), we use a square/cubic system of side $A=80$ and plate sizes in the range $L_{\min}=3$ to $L_{\max}=9$, with arrests imposed at sizes $5$, $6$, and $7$, respectively.

Figures~\ref{fig:2D_DSC}--\ref{fig:3DL_DSC} translate the size removal on reverse into a heat-flow surrogate $q(T)$ by mapping bins to monotonically ordered temperatures (Sec.~\ref{sec:model}). In all geometries, the A/C/E evolution (top rows) corresponds to a reproducible dip at a temperature close to the previously imposed arrest (bottom rows). The dip is shallowest for $2\mathrm{D}$, deeper for $3\mathrm{D}$, and sharpest for $3\mathrm{D}_{\!L}$, aligning with the degree of lateral jamming and the spectrum narrowing described above. Note also that after hammering (panels f) the dip becomes visibly narrower and deeper in all cases, even though its absolute position along the temperature axis remains essentially tied to the chosen arrest size.

The visual illustration presented in this Section will be complemented with some metrics we propose in the next Section.

\section{SMR and Entropy metrics across geometries}

As previously stated, an incomplete reverse transformation removes all plates with sizes smaller than or equal to $s$.
During the subsequent direct transformation, the system rebuilds a modified size distribution:
sizes $L > s$ repopulate from both surviving plates and new nucleation events,
while the bin at size $s$ develops a characteristic depletion caused by the geometric constraint imposed at arrest.
We quantify this imprint using two complementary metrics: the \emph{local} Size Mass Ratio (SMR) and the \emph{global} Shannon size--entropy $S$. These quantities are extracted directly from the same mass-per-size histograms that appear in Figs.~\ref{fig:Sizes2D}--\ref{fig:Sizes3DLam}, so the reader can always cross-check the numerical values against the visual trends.

\subsection{Local memory metric: Size Mass Ratio (SMR)}

To characterize the sharpness of the depletion at the arrested size $s$, we introduce the non-dimensional
\begin{equation}
\mathrm{SMR}(s)=\frac{M_{s-1}+M_{s+1}}{2\,M_s},
\label{eq:SMRdef}
\end{equation}
where $M_k$ is the ensemble-averaged mass contained in size bin $k$
(mass $\propto L^2$ in $2\mathrm{D}$, $L^3$ in $3\mathrm{D}$, and $L^2\!\times\!TW$ in $3\mathrm{D}_{\!L}$ for fixed lamellar thickness $TW$).
Values $\mathrm{SMR}(s)>1$ indicate that the arrested bin is locally depleted relative to its neighbors.

Table~\ref{tab:SMR_by_size_final} reports SMR$(s)$ for all geometries and arrest sizes.
As expected, the non-A instances (C and E) always satisfy SMR$(s)>1$, since the arrest bin becomes a local minimum of the size--mass spectrum.
The hammer step (E) consistently produces a stronger depletion than C.
Across geometries, the lamellar system ($3\mathrm{D}_{\!L}$) yields the smallest SMR$(s)$ at fixed $s$, consistent with its weaker memory after the first arrest--regrowth and with the shallower depletion observed in the DSC surrogate.
Only after hammering does its SMR become comparable to that of $2\mathrm{D}$ and $3\mathrm{D}$.

\begin{table}[ht]
\centering
\begin{tabular}{c|c|ccc}
\hline
system & instance & $L_A=5$ & $L_A=6$ & $L_A=7$ \\
\hline
$2\mathrm{D}$  & A & 0.96499 & 0.98544 & 1.00556 \\
$2\mathrm{D}$  & C & 1.67907 & 1.61744 & 1.67314 \\
$2\mathrm{D}$  & E & 2.52924 & 2.48333 & 2.44014 \\
\hline
$3\mathrm{D}$  & A & 0.98968 & 1.08359 & 0.89526 \\
$3\mathrm{D}$  & C & 1.68597 & 1.64948 & 1.65830 \\
$3\mathrm{D}$  & E & 2.20109 & 1.99310 & 2.03379 \\
\hline
$3\mathrm{D}_{\!L}$ & A & 1.01550 & 1.06703 & 0.96123 \\
$3\mathrm{D}_{\!L}$ & C & 1.56200 & 1.47398 & 1.43485 \\
$3\mathrm{D}_{\!L}$ & E & 2.20010 & 2.17049 & 2.07347 \\
\hline
\end{tabular}
\caption{Size Mass Ratio SMR$(s)=(M_{s-1}+M_{s+1})/(2M_s)$ by arrest size $s\in\{5,6,7\}$, using the dataset corresponding to that arrest. For non-A instances the arrest bin is always a local minimum, hence SMR$(s)>1$.}
\label{tab:SMR_by_size_final}
\end{table}

Physically, SMR$(s)$ measures the contrast at the arrested size: values above unity indicate that mass at size $s$ is depleted relative to its neighboring bins.
Because each repeated partial reverse cycle re-applies the same geometric constraint while preserving the larger plates, hammering deepens the depletion and sharpens the spectral imprint.
The A rows remain close to unity because the initial jammed states do not yet exhibit selective depletion.  The progression A$\to$C$\to$E in Table~\ref{tab:SMR_by_size_final} can thus be interpreted as a quantitative version of the increasingly deep notch that one sees by eye in Figs.~\ref{fig:Sizes2D}--\ref{fig:Sizes3DLam}.

\subsection{Global metric: Shannon configurational entropy}

To complement the local SMR measure, we evaluate the global configurational Shannon entropy
\begin{equation}
S \;=\; -\sum_{L} p_L \,\ln p_L,
\end{equation}
where $p_L$ is the mass probability carried by size $L$.
In $2\mathrm{D}$, $p_L=\frac{n_L L^2}{\sum_{K} n_K K^2}$;
in $3\mathrm{D}$, $p_L=\frac{n_L L^3}{\sum_{K} n_K K^3}$;
and in $3\mathrm{D}_{\!L}$, $p_L=\frac{n_L (L^2\,TW)}{\sum_{K} n_K (K^2\,TW)}$.
The entropy reaches its maximum for a perfectly uniform size distribution and
decreases as the spectrum becomes narrower or otherwise unbalanced. For the
entropy defined in Eq.~2, the maximum value corresponds to an equal mass
distribution across all size bins and is given by $\ln N$, where $N$ is the
total number of bins. In the numerical data presented here (and also in
Section~IV), we used sizes from $L_{\min}=3$ to $L_{\max}=9$, giving
$N=7$. Therefore, the maximum entropy is
\[
S_{\max} = \ln 7 \approx 1.94591\ldots
\]
This value is very close to the entropy observed in the $2\mathrm{D}$ system immediately
after initial jamming, consistent with its nearly uniform size distribution.

Table~\ref{tab:S_all} summarizes $S$ for all geometries and arrest sizes.

For the $2\mathrm{D}$ geometry, $S$ decreases systematically from A$\!\to$C$\!\to$E, reflecting the progressive loss of configurational diversity as small sizes are removed at each arrest and the spectrum narrows during regrowth. The initial size distribution, before any arrest, is quasi-uniform, close to the highest attainable entropy, so a post-arrest decrease is inevitable.

For the $3\mathrm{D}$ system, $S$ decreases only negligibly after the first arrest (remaining essentially constant). This is because the initial distribution is already somewhat biased. After the second arrest (hammer) the decrease of $S$ is clearer.

In contrast, the lamellar system ($3\mathrm{D}_{\!L}$) begins with a strongly biased jammed state (A), with entropy already low. The first arrest--regrowth cycle (C) therefore increases $S$, followed by a modest decline from C$\!\to$E after hammering, while still remaining at a higher entropy than in instance A.

\begin{table}[ht]
\centering
\begin{tabular}{c|c|cc|cc|cc}
\hline
system & instance & $S(L_A{=}5)$ & $e^{S}$ & $S(L_A{=}6)$ & $e^{S}$ & $S(L_A{=}7)$ & $e^{S}$ \\
\hline
$2\mathrm{D}$  & A & 1.94567 & 6.9983 & 1.94540 & 6.9964 & 1.94536 & 6.9962 \\
$2\mathrm{D}$  & C & 1.89987 & 6.6850 & 1.89306 & 6.6397 & 1.88997 & 6.6192 \\
$2\mathrm{D}$  & E & 1.84947 & 6.3564 & 1.82315 & 6.1913 & 1.80860 & 6.1019 \\
\hline\hline
$3\mathrm{D}$  & A & 1.92823 & 6.8773 & 1.92743 & 6.8718 & 1.93002 & 6.8896 \\
$3\mathrm{D}$  & C & 1.92197 & 6.8344 & 1.92103 & 6.8280 & 1.91420 & 6.7815 \\
$3\mathrm{D}$  & E & 1.89948 & 6.6824 & 1.89475 & 6.6509 & 1.87284 & 6.5067 \\
\hline\hline
$3\mathrm{D}_{\!L}$ & A & 1.86386 & 6.4486 & 1.86721 & 6.4702 & 1.86814 & 6.4762 \\
$3\mathrm{D}_{\!L}$ & C & 1.89768 & 6.6704 & 1.89684 & 6.6648 & 1.88418 & 6.5810 \\
$3\mathrm{D}_{\!L}$ & E & 1.87870 & 6.5450 & 1.88630 & 6.5949 & 1.87219 & 6.5025 \\
\hline
\end{tabular}
\caption{Shannon entropy $S=-\sum_i p_i\ln p_i$ and its exponential $e^{S}$ for all geometries ($2\mathrm{D}$, $3\mathrm{D}$, $3\mathrm{D}_{\!L}$), for instances A, C, and E, and arrest sizes $L_A\in\{5,6,7\}$. All probabilities $p_i$ are mass-normalized.}
\label{tab:S_all}
\end{table}

The exponential $e^{S}$ serves as an ``effective number'' of equally populated size bins.
Reductions in $S$ and $e^{S}$ reflect spectral narrowing.
Such narrowing occurs consistently for $2\mathrm{D}$ and negligibly for $3\mathrm{D}$ (after first arrest) as A$\to$C$\to$E.
In contrast, the lamellar $3\mathrm{D}_{\!L}$ system exhibits the ordering sequence C$>$E$>$A, reflecting its initially overconstrained jammed state. From the point of view of Fig.~\ref{fig:Sizes3DLam}, this means that the first arrest partly ``rebalances'' the initially bottom-heavy lamellar spectrum before the second arrest nudges it back toward a more depleted state.

\begin{figure}[H]
  \centering
  \includegraphics[width=0.45\linewidth]{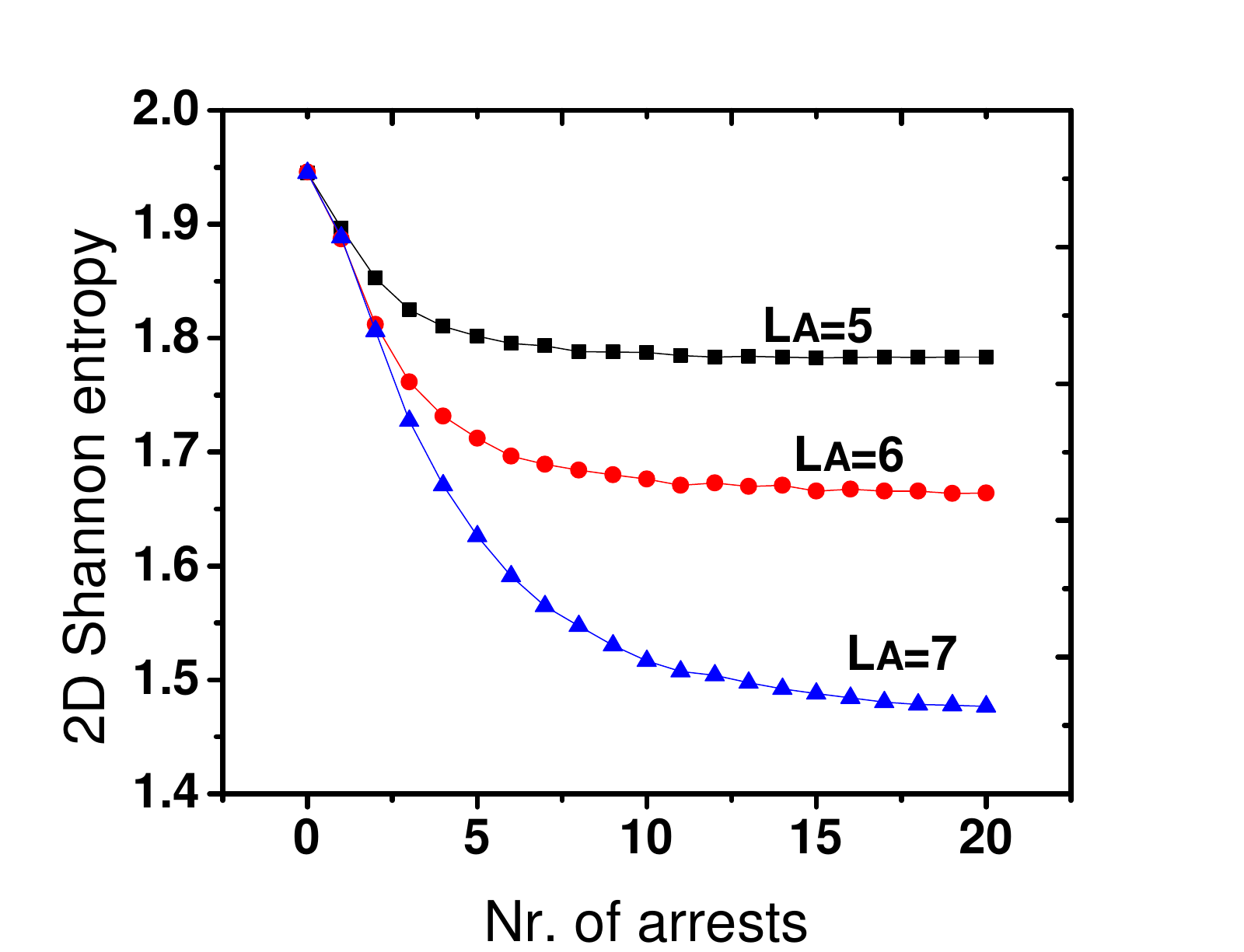}
\caption{Evolution of the Shannon entropy of the size distribution in the 2D system,
evaluated only at jammed configurations. The horizontal axis represents the
number of arrests (incomplete transformations with arrest sizes $L_A = 5,6,$
and $7$) that occurred before the corresponding jamming event. A value of ``0''
denotes the original jamming without previous arrests. Because each hammering
cycle consists of an arrest followed by regrowth, the final jam after 10
hammerings appears at $20$ on the horizontal axis.}
  \label{Entropy}
\end{figure}
The evolution of the configurational disorder is illustrated in Fig.~\ref{Entropy}, which shows the Shannon entropy of the size distribution evaluated only at jammed configurations. The horizontal axis represents the number of arrests (incomplete transformations) preceding each jamming event.
For all three arrest sizes considered ($L_A=5,6,$ and $7$), the entropy decreases monotonically with the number of arrests, indicating that repeated incomplete transformations progressively reduce the diversity of accessible plate sizes. Each arrest selectively removes smaller plates and constrains the geometry available for subsequent growth, leading to jammed states with increasingly restricted size distributions.
The rate of this entropy reduction depends on the arrest size. Smaller arrests (e.g., $L_A=5$) produce a faster approach toward saturation, whereas larger arrests (e.g., $L_A=7$) lead to a slower evolution, reflecting a weaker perturbation of the size distribution.
Because the simulations rely on a strict brute-force geometric growth algorithm, extending the full entropy-versus-hammering analysis to higher-dimensional systems would require very large computational effort while maintaining adequate ensemble statistics. For this reason, the systematic hammering dependence of the Shannon entropy is presented here for the 2D case, where the trend can be established clearly. Corresponding higher-dimensional calculations are left for future work.

We stress that the Shannon entropy calculated here is a ``mesoscopic'' one, referring to the size distribution of plates that are far larger than inter-atomic distances. This kind of entropy is useful for describing the diversity of plate sizes, but it should be thermodynamically irrelevant --- many orders of magnitude smaller --- when discussing the entropy change at the martensitic phase transition.

\section{Conclusions}

We introduced and analyzed a minimal geometry--driven phase--transition model, implemented in \(2\mathrm{D}\), \(3\mathrm{D}\), and \(3\mathrm{D}_{\!L}\), a lamellar variant. The forward transformation is modeled by random nucleation and finite self--similar (homothetic) growth of plates (squares, cubes, or lamellae, respectively), while the reverse transformation is non--random and is taken to proceed in reverse size order, with the smallest plates disappearing first. Consequently, an incomplete reverse transformation retains only the largest plates. These surviving plates dominate the subsequent direct transformation and, together with the geometrical constraints they impose, generate a depletion of intermediate sizes: this constitutes the memory effect analyzed in this paper.

If the same arrest cycle is imposed repeatedly (``hammering''), the geometrical restrictions become more pronounced. As a result, the selective depletion becomes sharper, producing a deeper and narrower minimum in the recovered size distribution. In our simulations, this sharpening manifests as an increasingly pronounced dip in the DSC-like synthetic signal, faithfully reproducing the experimental memory effect in shape memory alloys.

Before any arrest, the initial jamming produces a quasi-even size--mass distribution in $2\mathrm{D}$, a slightly biased one in $3\mathrm{D}$, and a strongly biased distribution in $3\mathrm{D}_{\!L}$, favoring an increased number of smaller plates. After the arrest(s), the modification of the plate-size distribution is {\it visually} clear and significant in all dimensionalities, both in the averaged bar charts and in the simulated DSC signals. It is the \emph{quantification} we propose that reveals the differences in the memory capacities across the dimensionalities, and our model suggests that the memory effect is overall stronger in $2\mathrm{D}$ than in $3\mathrm{D}$ and $3\mathrm{D}_{\!L}$.

To measure how strongly an arrest leaves a trace in the microstructure, we used
two complementary indicators. The first is the \emph{size--mass ratio} (SMR), a
local measure that compares the population of the arrest-size bin with its
neighbors. Initial distributions give $\mathrm{SMR}\approx 1$, whereas
$\mathrm{SMR}>1$ signals a memory imprint. The second is the \emph{Shannon
size--entropy}, a global metric that reflects how balanced or skewed the entire
size--mass distribution is.

Taken together, these measures show a consistent hierarchy. After the first
arrest, the $2\mathrm{D}$ and $3\mathrm{D}$ systems display comparable SMR
values, while the lamellar geometry ($3\mathrm{D}_{\!L}$) shows a weaker imprint
because its initial distribution is already strongly biased toward small plates.
After the second arrest, the SMR in $3\mathrm{D}$ and $3\mathrm{D}_{\!L}$ becomes
similar, but both remain significantly below the $2\mathrm{D}$ response, for which
our data suggest the strongest memory effect. The entropy trends mirror this
ordering: the largest entropy drop occurs in $2\mathrm{D}$, which starts from an
almost uniform distribution; the entropy in $3\mathrm{D}$ stays nearly unchanged
after the first arrest; and $3\mathrm{D}_{\!L}$ even shows a slight increase,
reflecting its already low-entropy, highly skewed initial state. For the $2\mathrm{D}$ case, where long hammering sequences can be explored systematically, we extended the protocol up to 20 cycles and found that the Shannon entropy of the size distribution progressively decreases and eventually reaches a saturation (plateau). This saturation occurs more rapidly when the arrest is imposed at smaller sizes.

Finally, our model conclusively captures the memory effect in incomplete transformations using geometry alone, without relying on explicit thermodynamic assumptions. This isolation of geometric mechanisms---finite growth and domain blocking---allows their influence to be examined cleanly and systematically across all dimensionalities considered.

{\bf Outlook:} We hope that our paper motivates further research on the TME, which in practice is more easily observed in ribbons (effectively closer to $2\mathrm{D}$) than in bulk ($3\mathrm{D}$), consistent with the trends predicted here. Beyond the thermal memory itself, our study suggests that incomplete transformations can induce significant microstructural changes, and future work may examine how these geometry-induced memory imprints interact with additional physical ingredients (e.g., elastic or magnetic energies) and whether such coupling produces measurable macroscopic consequences.

\vskip 1cm

{\bf Acknowledgements.}
This work was supported by the Romanian Ministry of Research, Innovation and Digitization, Core Program Project grant number PC2-PN23080202.

\vskip 1cm

\noindent{\bf Use of AI tools.}
ChatGPT was used for minor language editing and for assistance in phrasing the extension from our earlier $2\mathrm{D}$ mechanism to the $3\mathrm{D}$ and $3\mathrm{D}_{\!L}$ variants. All scientific ideas, simulations, analyses, figures, and conclusions are the authors’ own. All AI-suggested text was reviewed and revised by the authors, who take full responsibility for the content of this work.

\vskip 1cm

{\bf Declaration of competing interest.}
The authors declare that they have no competing financial interests.


\begin{thebibliography}{99}

\bibitem{Landau1937}
L.~D.~Landau,
``On the Theory of Phase Transitions,''
Zh.\ Eksp.\ Teor.\ Fiz.\ \textbf{7}, 19 (1937).

\bibitem{LLStatPhys}
L.~D.~Landau and E.~M.~Lifshitz,
\emph{Statistical Physics, Part I}, 3rd ed.
(Pergamon, Oxford, 1980).

\bibitem{CahnHilliard1958}
J.~W.~Cahn and J.~E.~Hilliard,
``Free Energy of a Nonuniform System. I. Interfacial Free Energy,''
J.\ Chem.\ Phys.\ \textbf{28}, 258 (1958).

\bibitem{Khachaturyan1983}
A.~G.~Khachaturyan,
\emph{Theory of Structural Transformations in Solids}
(Wiley, New York, 1983).

\bibitem{Avrami1939}
M.~Avrami,
``Kinetics of Phase Change. I,''
J.\ Chem.\ Phys.\ \textbf{7}, 1103 (1939);
``Kinetics of Phase Change. II,''
J.\ Chem.\ Phys.\ \textbf{8}, 212 (1940);
``Kinetics of Phase Change. III,''
J.\ Chem.\ Phys.\ \textbf{9}, 177 (1941).

\bibitem{Binder1987}
K.~Binder,
``Theory of First-Order Phase Transitions,''
Rep.\ Prog.\ Phys.\ \textbf{50}, 783 (1987).

\bibitem{Krumhansl1989}
J.~A.~Krumhansl and J.~R.~Gooding,
``Structural Phase Transitions with Pseudospin Variables: A Microscopic Theory,''
Phys.\ Rev.\ B \textbf{39}, 3047 (1989).

\bibitem{Lookman2003}
T.~Lookman, S.~R.~Shenoy, K.~O.~Rasmussen, A.~Saxena, and A.~R.~Bishop,
``Ferroelastic Dynamics and the Role of Microstructure,''
Phys.\ Rev.\ B \textbf{67}, 024114 (2003).

\bibitem{Otsuka}
T.~Tadaki, K.~Otsuka, and K.~Shimizu,
``Shape Memory Alloys,''
Annu.\ Rev.\ Mater.\ Sci.\ \textbf{18}, 25 (1988).

\bibitem{Planes}
E.~Bonnot, R.~Romero, L.~Ma\~nosa, E.~Vives, and A.~Planes,
``Elastocaloric Effect Associated with the Martensitic Transition in Shape-Memory Alloys,''
Phys.\ Rev.\ Lett.\ \textbf{100}, 125901 (2008).

\bibitem{Jani}
J.~M.~Jani, M.~Leary, A.~Subic, and M.~A.~Gibson,
``A Review of Shape Memory Alloy Research, Applications and Opportunities,''
Mater.\ Des.\ \textbf{56}, 1078 (2014).

\bibitem{Vasilev}
A.~N.~Vasilev, A.~D.~Bozhko, V.~V.~Khovailo, I.~E.~Dikshtein, V.~G.~Shavrov,
V.~D.~Buchelnikov, M.~Matsumoto, S.~Suzuki, T.~Takagi, and J.~Tani,
``Structural and Magnetic Phase Transitions in Shape-Memory Alloys $\mathrm{Ni_{2+x}Mn_{1-x}Ga}$,''
Phys.\ Rev.\ B \textbf{59}, 1113 (1999).

\bibitem{Pasquale}
M.~Pasquale, C.~P.~Sasso, L.~H.~Lewis, L.~Giudici, T.~Lograsso, and D.~Schlagel,
``Magnetostructural Transition and Magnetocaloric Effect in $\mathrm{Ni_{55}Mn_{20}Ga_{25}}$ Single Crystals,''
Phys.\ Rev.\ B \textbf{72}, 094435 (2005).

\bibitem{Kaufmann}
S.~Kaufmann, U.~K.~Rößler, O.~Heczko, M.~Wuttig, J.~Buschbeck,
L.~Schultz, and S.~Fähler,
``Adaptive Modulations of Martensites,''
Phys.\ Rev.\ Lett.\ \textbf{104}, 145702 (2010).

\bibitem{Zheludev}
A.~Zheludev, S.~M.~Shapiro, P.~Wochner, A.~Schwartz, M.~Wall, and L.~E.~Tanner,
``Phonon Anomaly, Central Peak, and Microstructures in Ni$_2$MnGa,''
Phys.\ Rev.\ B \textbf{51}, 11310 (1995).

\bibitem{Umetsu}
R.~Y.~Umetsu, X.~Xu, and R.~Kainuma,
``NiMn-Based Metamagnetic Shape Memory Alloys,''
Scripta Mater.\ \textbf{116}, 1 (2016).

\bibitem{Nnamchi}
P.~Nnamchi, A.~Younes, and S.~González,
``A Review on Shape Memory Metallic Alloys and Their Critical Stress for Twinning,''
Intermetallics \textbf{105}, 61 (2019).

\bibitem{Tolea1}
F.~\c{T}olea, M.~\c{T}olea, M.~Sofronie, and M.~Valeanu,
``Distribution of Plates' Sizes Tell the Thermal History in a Simulated Martensitic-Like Phase Transition,''
Solid State Commun.\ \textbf{213--214}, 37 (2015).

\bibitem{ToleaPRE}
F.~\c{T}olea, M.~Sofronie, M.~Ni\c{t}\u{a}, and M.~\c{T}olea,
``Memory of Incomplete Phase Transitions from a Random Squares Model,''
Phys.\ Rev.\ E \textbf{108}, 064134 (2023).

\bibitem{Airoldi1}
G.~Airoldi, A.~Corsi, and R.~Riva,
``The Hysteresis Cycle Modification in Thermoelastic Martensitic Transformation of Shape Memory Alloys,''
Scripta Mater.\ \textbf{36}, 1273 (1997).

\bibitem{Mad-Scripta2}
K.~Madangopal,
``New Observations on the Thermal Arrest Memory Effect in Ni--Ti Alloys,''
Scripta Mater.\ \textbf{53}, 875 (2005).

\bibitem{R-AACTA2}
J.~Rodriguez-Aseguinolaza, I.~Ruiz-Larrea, M.~L.~No, A.~Lopez-Echarri, and J.~San~Juan,
``Temperature Memory Effect in Cu--Al--Ni Shape Memory Alloys Studied by Adiabatic Calorimetry,''
Acta Mater.\ \textbf{56}, 3711 (2008).

\bibitem{Cui}
Y.~Zheng, L.~Cui, and J.~Schrooten,
``Temperature Memory Effect of a Nickel--Titanium Shape Memory Alloy,''
Appl.\ Phys.\ Lett.\ \textbf{84}, 31 (2004).

\bibitem{Liu}
T.~Liu, Y.~Zheng, and L.~Cui,
``Influence of Partial Cycling on the Transformation Mass of NiTi Alloys,''
Mater.\ Lett.\ \textbf{112}, 121 (2013).

\bibitem{Tang}
C.~Tang, T.~X.~Wang, W.~M.~Huang, L.~Sun, and X.~Y.~Gao,
``Temperature Sensors Based on the Temperature Memory Effect in Shape Memory Alloys to Check Minor Over-Heating,''
Sens.\ Actuators A \textbf{238}, 337 (2016).

\bibitem{WANG_SM}
Z.~G.~Wang and X.~T.~Zu,
``Incomplete Transformation Induced Multiple-Step Transformation in TiNi Shape Memory Alloys,''
Scripta Mater.\ \textbf{53}, 335 (2005).

\bibitem{Tolea2}
F.~\c{T}olea, M.~\c{T}olea, and M.~Valeanu,
``Thermal Memory Fading by Heating to a Lower Temperature: Experimental Data on Polycrystalline NiFeGa Ribbons and 2D Statistical Model Predictions,''
Solid State Commun.\ \textbf{257}, 36 (2017).

\bibitem{Vidal2}
A.~Vidal-Crespo, A.~F.~Manchón-Gordón, J.~M.~Martín-Olalla, F.~J.~Romero,
J.~J.~Ipus, M.~C.~Gallardo, and J.~S.~Blázquez,
``Phase Dependence of the Thermal Memory Effect in Polycrystalline Ribbon and Bulk Ni$_{55}$Fe$_{19}$Ga$_{26}$ Heusler Alloys,''
Intermetallics \textbf{180}, 108695 (2025).

\bibitem{ToleaJALCOM}
F.~\c{T}olea, M.~Ni\c{t}\u{a}, and M.~\c{T}olea,
``Thermal Memory Effect in NiFeGa and NiMnGa Shape Memory Ribbons: Toward Maximum-Temperature Recording Applications,''
J.\ Alloys Compd.\ \textbf{1043}, 184056 (2025).

\bibitem{VidalCrespoTA}
A.~Vidal-Crespo, A.~F.~Manch\'on-Gord\'on, J.~J.~Ipus, and J.~S.~Bl\'azquez,
``Thermal memory effect in Mn(CoFe)Ge intermetallic compound,''
Thermochim.\ Acta \textbf{756}, 180200 (2026).

\bibitem{Rao}
M.~Rao and S.~Sengupta,
``Droplet Fluctuations in the Morphology and Kinetics of Martensites,''
Phys.\ Rev.\ Lett.\ \textbf{78}, 2168 (1997).

\bibitem{Perez}
F.~J.~Pérez-Reche, E.~Vives, L.~Ma\~nosa, and A.~Planes,
``Athermal Character of Structural Phase Transitions,''
Phys.\ Rev.\ Lett.\ \textbf{87}, 195701 (2001).

\bibitem{Recarte}
J.~I.~Perez-Landazabal, V.~Recarte, V.~Sanchez-Alarcos, S.~Kustov,
D.~Salas, and E.~Cesari,
``Effect of Magnetic Field on the Isothermal Transformation of a Ni--Mn--In--Co Magnetic Shape Memory Alloy,''
Intermetallics \textbf{28}, 144 (2012).

\bibitem{Samaniego}
E.~Samaniego, C.~Anitescu, S.~Goswami, V.~M.~Nguyen-Thanh, H.~Guo,
K.~Hamdia, X.~Zhuang, and T.~Rabczuk,
``An energy approach to the solution of partial differential equations in computational mechanics via machine learning: Concepts, implementation and applications,''
Comput. Methods Appl. Mech. Engrg. \textbf{362}, 112790 (2020).

\end{thebibliography}
\end{document}